\documentclass[11pt,a4paper,onecolumn,final]{article}
\usepackage{srcltx}
\usepackage{graphicx}
\usepackage{amsmath}
\usepackage{natbib}
    \renewcommand{\cite}{\citep}
\bibpunct{(}{)}{;}{a}{}{,}
\begin{document}
\title{Detection of residual native state entropy changes upon mutation in Fyn SH3}
\author{%
    Kresten Lindorff-Larsen$^{1,2,3}$, %
    Robert B. Best$^{1,4}$, %
    Anthony Mittermaier$^{5,6}$,\\%
    Lewis E. Kay$^{5,7}$, %
    Christopher M. Dobson$^{1,8}$, %
    and Michele Vendruscolo$^{1,\ast}$
    }%

\date{}
\maketitle%
$^{1}$Department of Chemistry, University of Cambridge, %
Lensfield Road, Cambridge, CB2 1EW, United Kingdom%

$^{2}$Department of Biochemistry, Institute of Molecular Biology and Physiology, %
University of Copenhagen, Universitetsparken 13, DK-2100 Copenhagen {\O}, Denmark%

$^{3}$Current address: Structural Biology and NMR Laboratory \& the Linderstr{\o}m-Lang Centre for Protein Science, Department of Biology, University of Copenhagen, 2200 Copenhagen, Denmark

$^{4}$Laboratory of Chemical Physics, NIDDK, %
National Institutes of Health, Bethesda, Maryland 20892-0520,
U.S.A.%

$^{5}$Department of Biochemistry, University of Toronto, Toronto, Ontario,
Canada M5S 1A8

$^{6}$Current address: Department of Chemistry, McGill University, 801 Sherbrooke St. W., Montreal, QC H3A 0B8, Canada

$^{7}$Departments of Medical
Genetics and Chemistry, University of Toronto, Toronto, Ontario,
Canada M5S 1A8

$^{8}$Deceased

$^{\ast}$Corresponding author. Email: mv245@cam.ac.uk

\newpage
%
%
\section*{Abstract\footnote{This manuscript was written in Cambridge, Toronto and Copenhagen in 2004--2005, forms part of K.L.-L.'s PhD thesis \cite{lindorff04}, and was uploaded to arXiv in 2026.}%
}
\textbf{%
NMR relaxation experiments have shown that there are small but
measurable changes in the native state dynamics of the Fyn SH3
domain associated with the substitution by other amino acids of a
phenylalanine residue (F20) in the hydrophobic core. We have here
used experimental values of NMR order parameters for the wild type
protein and two mutational variants (F20L and F20V) as restraints in
molecular dynamics simulations. This approach is highly sensitive
and provides an atomistic description of the subtle perturbations in
native state fluctuations accompanying the mutations. The structural
ensembles that we have determined using this method allow the
changes in the native state entropy of the protein caused by each of
the mutations to be estimated. These entropy changes correspond to
free energy variations of several kcal mol$^{-1}$ and therefore
represent sizable contributions to the overall changes in stability
that are associated with the amino acid mutations.
}%
\\[1cm]%
\textbf{Keywords}: Entropy/Mutation/NMR relaxation/Protein
dynamics/Protein stability
\newpage
%
%
\section*{Introduction}
Natural mutations in proteins are at the heart of molecular
evolution \cite{graur00}, and are the origins of many familial
diseases \cite{dobson03}. In the laboratory, the introduction of
specific mutations through the technique of protein engineering
represents a powerful approach that can be used to perturb in a
controlled manner a wide range of properties of a protein.
Consequently, the effects of amino acid mutations on function
\cite{leatherbarrow86,brannigan02}, structure
\cite{eriksson92,baldwin93a,buckle93}, stability
\cite{serrano89,serrano92,carter01,gromiha02,guerois02,kortemme02,bordner04,capriotti05},
folding \cite{evans87,matouschek89,fersht92} and aggregation
\cite{chiti00,chiti02,chiti03,delapaz04,dubay04,pawar05} have been
the focus of much research.

The dynamical aspects of proteins are now recognized as playing a
central role in their function
\cite{mccammon77,frauenfelder91,rasmussen92,eisenmesser02,karplus02,wong03,benkovic03}.
Further, protein dynamics is intimately linked to the thermodynamics
of protein stability and ligand affinity
\cite{cooper76,akke93,yang96a,li96a,zidek99,wrabl00,mittermaier04,spyracopoulos05}.
Protein engineering methods have the possibility  to provide unique
insights into the determinants of protein mobility, and recently the
details of the changes in native state dynamics that occur upon
amino acid mutations have been the subject of a variety of
experimental studies
\cite{stone01,stone01a,mayer03,millet03,mittermaier04}.

In this paper we describe an approach to perform a structural
analysis of the ensembles of protein conformations that are
populated in the native state and to determine thermodynamic
consequences of the changes  in native state dynamics that accompany
amino acid substitutions. In the technique that we use, data from
heteronuclear NMR relaxation experiments are combined with computer
simulations in order to describe native state heterogeneity. In
particular, experimentally derived backbone and side-chain order
parameters for the Fyn SH3 domain and two hydrophobic core mutants
(F20L and F20V) \cite{mittermaier04} are used as restraints in
molecular dynamics (MD) simulations \cite{best04,lindorff05a}. In
these simulations a molecular mechanics force-field is used as a
motional model to interpret experimental NMR relaxation data for the
two mutants, and thus to provide a structural description of the
changes in native state heterogeneity associated with the alteration
of the fluctuations that occur upon amino acid substitution. Our
results show that both mutations cause small, but measurable,
changes in such heterogeneity. Although the native states of the
F20L and F20V mutational variants are destabilized relative to the
unfolded state in terms of the total free energy, the increased
internal mobility caused by the two mutations corresponds to an
entropic stabilization of the native state of several kcal
mol$^{-1}$. This observation has important implications for the
interpretation of protein engineering data and for understanding the
factors that control protein stability.
%
\section*{Methods}
\subsection*{Molecular dynamics simulations}
Molecular dynamics simulations of chicken Fyn SH3 and two of its
mutational variants (F20L and F20V) were performed using CHARMM
\cite{brooks83}. The structure of chicken Fyn SH3 has not been
determined experimentally but is known to be very similar to that of
the human variant \cite{mittermaier04}. Therefore, a model of the
structure of the chicken isoform, herein termed Fyn SH3, was
generated by introducing the two mutations V1S and V5E into the
structure of the human isoform (PDB entry 1SHF) using Deep-View
\cite{schwede03} and subsequently energy minimizing the resulting
structure. Also, the two mutants of Fyn SH3, F20L and F20V, are
known to have a virtually identical backbone structure to that of
the wild-type (WT) protein \cite{mittermaier04}; these two mutations
were therefore modelled in a similar manner. Simulations were
carried out using a polar hydrogen model and Param19 force-field
\cite{neria96}. The effects of water were modelled by surrounding
the proteins by a 5{\AA} solvation shell consisting of 724 TIP3P
water molecules \cite{jorgensen83}; a boundary potential was used to
prevent water molecules from escaping \cite{beglov95}. All
simulations were carried out at 298~K.

Experimentally determined squared order parameters ($S^2$ values)
were used as restraints \cite{best04}. In this type of simulations a
set of molecules is simulated in parallel since the order parameter
for a single molecule at any given time is by definition unity. It
has previously been shown that simulations using 8, 16 or 32
molecules (replicas) provide very similar and well converged results
\cite{best04}. All simulations reported here were obtained by
simulating an ensemble consisting of $N_{rep}=16$ replicas of the
protein.

At each step of the MD simulations $S^2$ values were calculated
across this ensemble using \cite{henry85,best04}
\begin{equation}
S^2_{ij}=\frac{3}{2}(%
    \langle \hat{x}^2_{ij}\rangle^2+\langle \hat{y}^2_{ij}\rangle^2+\langle \hat{z}^2_{ij}\rangle^2%
    +2\langle \hat{x}_{ij}\hat{y}_{ij}\rangle^2%
    +2\langle \hat{x}_{ij}\hat{z}_{ij}\rangle^2%
    +2\langle \hat{y}_{ij}\hat{z}_{ij}\rangle^2%
    )-\frac{1}{2}%
\end{equation}
where $S^2_{ij}$ is the calculated squared order parameter between
atoms $i$ and $j$, and $\hat{x}_{ij}$, $\hat{y}_{ij}$ and
$\hat{z}_{ij}$ are the components of a unit vector along this bond.
The resulting $S^2_{calc}$ were then compared to the $N$
experimentally determined values according to
\begin{equation}
\rho(t)=\frac{1}{N}\sum_{k=1}^{N}(S^2_{k,calc}(t)-S^2_{k,exp})^2
\end{equation}
The value of $\rho(t)$ therefore measures the mean square deviation
between experimental and calculated $S^2$ values. During the
simulations $\rho$ was progressively decreased by performing biased
MD using the energetic penalty \cite{paci99,best04}
\begin{equation}
    \label{eq:bmd}
    E(\rho(t)) = %
        \begin{cases}
            \frac{\alpha N_{rep}}{2}(\rho(t)-\rho_0(t))^2    & \text{if $\rho(t)>\rho_0(t)$} \\%
            0                                                & \text{if $\rho(t)\le \rho_0(t)$}%
        \end{cases}
\end{equation}
where $\alpha$ is a force-constant that determines the weight of the
$S^2$ restraints relative to the molecular mechanics force field,
and $\rho_0(t)$ is defined as
\begin{equation}
\rho_0(t)=\underset{0\le\tau\le t}{\min} \rho(\tau)
\end{equation}

The simulations were initiated with a 200~ps heating phase with
$\alpha=10^3$~kcal mol$^{-1}$. Then, over a simulation period of
140~ps, $\alpha$ was progressively increased to $1.6\cdot10^7$~kcal
mol$^{-1}$. Finally a 1.0 ns production phase
($\alpha=1.0\cdot10^7$~kcal mol$^{-1}$) was used to sample the
native state variability of Fyn SH3 and the two mutational variants.
The convergence of these simulations was estimated by splitting the
simulation up into four equally sized parts and analysing each
separately. The quasiharmonic analysis of the configurational
entropy (see below) revealed that fully converged values could be
obtained from the last half of the simulations. Thus, by saving
structures every 10 ps we obtained ensembles consisting of 800
structures (50 from the last 500ps for each replica) for each of the
three proteins.

A total of 44 (17), 48 (22) and 44 (19) amide (methyl) $S^2$
restraints were used for WT, F20L and F20V, respectively, after
discarding experimental $S^2$ values that may contain contributions
from slower motions on the ns timescale (amide groups of E5, S31,
S32, E33, G34, A39 and methyl groups of L42, V58)
\cite{mittermaier04}. All data were obtained at 298~K, the same
temperature as that used in the simulations.
\subsection*{Quasi-harmonic analysis}
A quasi-harmonic analysis \cite{karplus81a,levy84,case94,brooks95}
was performed in CHARMM. The vibrational contribution to the native
state entropy ($S_{vib}$) was estimated from the frequencies
obtained \cite{karplus81a,levy84,brooks95,mcquarrie00}. Before the
analysis all water molecules and hydrogen atoms were removed.
Reported numbers correspond to $T\Delta S_{vib}$. Error estimates of
$S_{vib}$ were obtained by a bootstrap procedure \cite{efron86}. One
hundred ensembles, each containing 800 structures, were sampled
(with replacement) from the original ensemble, and each ensemble was
subsequently used as input to a quasiharmonic analysis. The reported
error bounds are the standard deviations across these 100 ensembles.
As noted above, convergence was evaluated by analysing the total
simulation in equally sized blocks, and the values reported are
fully converged within the error bounds shown.
%
\section*{Results}
\subsection*{Ensembles representing the native state dynamics}
\begin{figure}[h!]
      \centering
      \includegraphics[width=12.0cm]{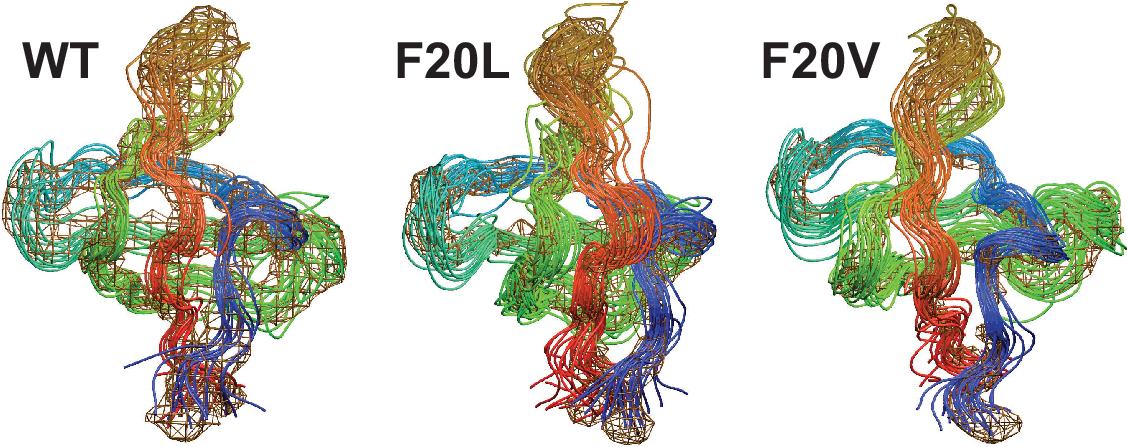}
      \caption{\label{fig:structures}%
      }
\end{figure}

\begin{figure}[h!]
      \centering
      \includegraphics[width=12.0cm]{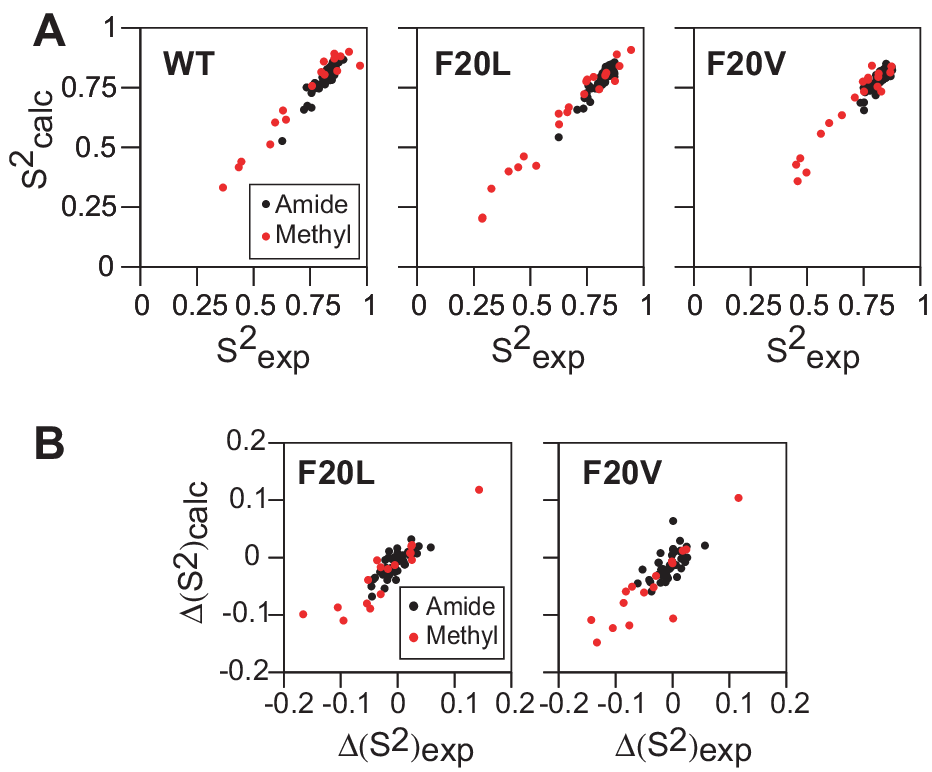}
      \caption{\label{fig:order-parameters-delta}%
      }
\end{figure}

Order parameters ($S^2$ values) derived from NMR relaxation
experiments provide a high resolution measure of the amplitudes of
motions in proteins \cite{kay98,kay05}. Thus, despite the fact that
the ns timescale fluctuations in WT Fyn SH3 and in the two mutants
(F20L and F20V) are very similar, the experimental data can be
obtained with sufficient accuracy \cite{millet02a} to detect small
local differences \cite{mittermaier04}. If these subtle differences
are to be interpreted in structural terms, it is therefore crucial
that the structural models reproduce the experimental data with high
precision. It is extremely difficult to obtain this level of
accuracy by a direct integration of the equation of motions using
conventional force-fields \cite{best04b}. We therefore carried out a
series of MD simulations in which the experimental amide and
side-chain $S^2$ values were used directly as restraints
\cite{best04,lindorff05a}, thereby introducing the experimentally
observed variability into the structural ensembles
(Figure~\ref{fig:structures}). The agreement between experimental
and calculated $S^2$ values is shown in
Figure~\ref{fig:order-parameters-delta}A; the correlations and root
mean square deviations (RMSDs) between the two sets of $S^2$ values
are $r^2=0.93$ ($\textrm{RMSD}=0.038$), 0.96 (0.043) and 0.91
(0.043) for WT, F20L and F20V, respectively, showing the very good
agreement that can be obtained between the experimental and
predicted $S^2$ values when the latter are imposed as restraints.
Nevertheless, despite these high correlations, it remains important
to examine the level to which the changes in $S^2$ values between
the WT and mutant proteins are reproduced by the calculated
structures (Figure~\ref{fig:order-parameters-delta}B). There is a
very good agreement, with correlations and RMSDs between
experimental and calculated values of $\Delta(S^2)$ of $r^2=0.79$
and $\textrm{RMSD}=0.020$ for F20L and $r^2=0.71$ and
$\textrm{RMSD}=0.025$ for F20V. The $q$-factors, defined as
$q=\sqrt{\sum( \Delta S^2_{calc}-\Delta S^2_{exp})^2}/\sqrt{\sum
(\Delta S^2_{calc})^2}$, are 0.51 for both F20L and F20V.

\begin{figure}[h!]
      \centering
      \includegraphics[width=12.0cm]{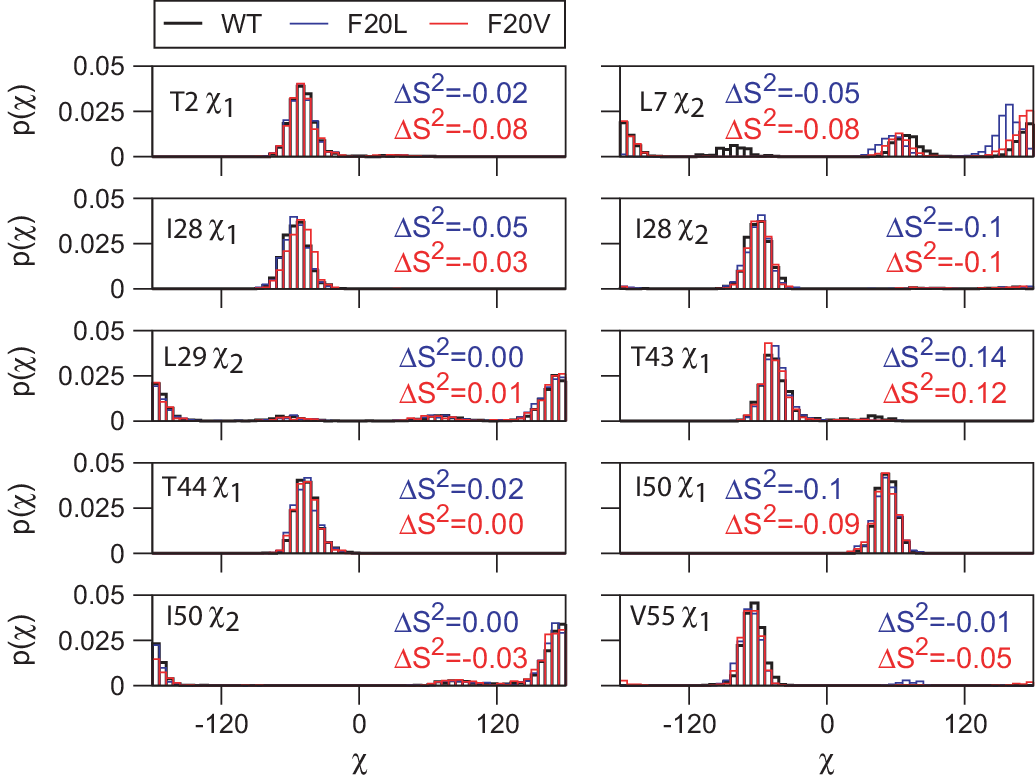}
      \caption{\label{fig:dihedrals}%
      }
\end{figure}

Given this high level of agreement between experimental and
calculated $S^2$ values, it becomes possible to provide a detailed
analysis of the motions associated with the observed order
parameters and their changes upon mutation. As an example, we show
the distribution of ten side-chain dihedral angles in the WT and
mutant proteins (Figure~\ref{fig:dihedrals}). It is clear that the
molecular motion giving rise to $S^2<1$ can include contributions
from both variability within a single energy well (e.g. `diffusion
in a cone'-type motion \cite{kay98}) and from transitions between
wells (rotameric transitions). No simple motional models will
therefore be able to describe well the overall fluctuations in the
structure. The high similarity between the dihedral angle
distributions obtained for the WT and mutant proteins is a
consequence of the very similar experimental $S^2$ values for the
three variants, and in addition shows that the MD simulations have
converged well. These results indicate that the experimental changes
in side-chain $S^2$ values result mainly from subtle shifts in the
populations of side-chain rotamers associated with the amino acid
substitutions at position 20 in the Fyn SH3 domain.
\subsection*{Mutational changes to the residual native state entropy}
It has been suggested that the mobility probed by experimental $S^2$
values is associated with significant entropy contributions
\cite{akke93,yang96a,li96a,wrabl00,spyracopoulos05}. A quantitative
analysis of this effect is complicated by the theoretical and
practical difficulties in estimating absolute values of entropies.
Nevertheless, using a range of motional models for amide and methyl
groups, an analytic relationship has been derived relating changes
in order parameters to entropy changes \cite{yang96a}. Based on this
method and the experimental amide and methyl $S^2$ values in Fyn SH3
it has been estimated that the F20L and F20V mutations are
accompanied by an entropic stabilization ($T\Delta S_{vib}$) of the
native state by 1.95 and 3.82 kcal mol$^{-1}$, respectively
\cite{mittermaier04}. Notably, these values are comparable in
magnitude to the free energy of unfolding of the WT Fyn SH3 domain
(-2.4~kcal mol$^{-1}$) \cite{northey02}. Overall, however, the F20L
and F20V mutations destabilize the protein relative to WT Fyn SH3 by
1.1 and 1.9 kcal mol$^{-1}$, respectively \cite{northey02}. These
estimates of the entropic contributions to stability changes
represent, therefore, a large fraction of the net free energy of the
native protein relative to its denatured state. As the values for
the entropy terms are based on simplified models for the motion
associated with the experimental $S^2$ values, it is important to
validate them by alternative methods of analysis. In addition, these
estimates are based on the assumption that the overall entropy can
be obtained as the sum of individual contributions, which is
strictly valid only if all intramolecular motions are independent of
each other; for correlated motions, this assumption will lead to
overestimates of the entropic contributions to the native state
stability.

\begin{figure}[h!]
      \centering
      \includegraphics[width=8.0cm]{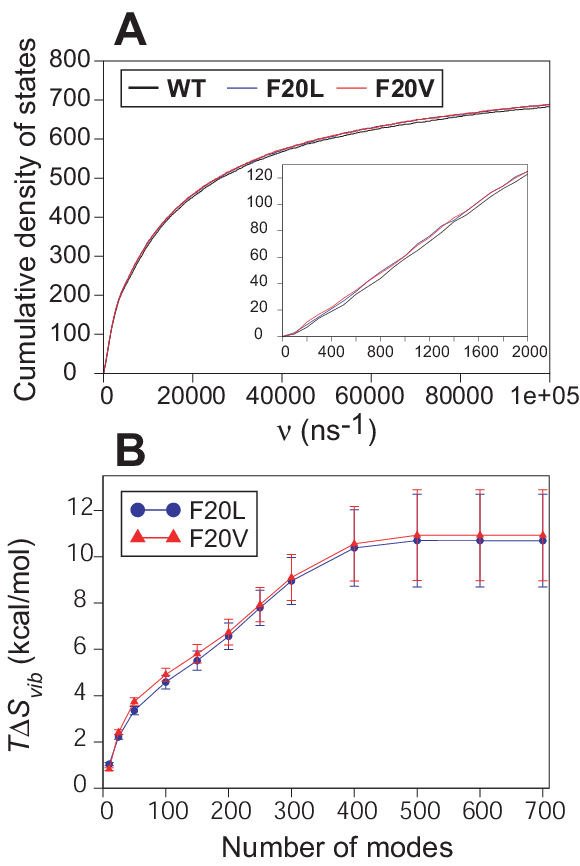}
      \caption{\label{fig:quasi}%
      }
\end{figure}

In the light of these considerations, we performed a quasi-harmonic
analysis \cite{brooks95} of the structural ensembles obtained in
this study to determine the density of states of vibrational modes
in the three variants (Figure~\ref{fig:quasi}A). The analysis was
carried out for the side chains for which experimental order
parameters were available in all three proteins as well as for the
polypeptide backbone. The results show that both F20L and F20V
display more extensive low frequency motions than the WT protein, in
particular in the range $\nu=\textrm{50--1000\,ns}^{-1}$. From the
frequencies involved it is possible to estimate the vibrational
contribution to the native state entropy
\cite{karplus81a,levy84,brooks95}. The results are shown in
Figure~\ref{fig:quasi}B and show that the changes in vibrational
entropy upon mutation correspond to large changes in free energy
($T\Delta S_{vib}$). Using, for example, the first 250 modes we
obtain estimates of $7.8\pm0.8$ and $7.9\pm0.7\textrm{kcal
mol}^{-1}$ for F20L and F20V, respectively.
\section*{Discussion}
Protein engineering is an important method of modulating protein
stability and many studies have focused on the rationalization and
prediction of changes in the native state stability accompanying
amino acid substitutions
\cite{serrano92,doig95,carter01,gromiha02,guerois02,kortemme02,bordner04}.
These methods are based on the assumption that the changes in
stability resulting from a mutation can be related to changes in
intramolecular interaction energies and solvation energies, and in
the configurational entropy of the unfolded protein; changes in the
residual native state entropy are often assumed to be negligible.
NMR relaxation measurements, however, provide evidence that there
are often measurable changes in native state dynamics associated
with mutations
\cite{stone01a,millet03,mayer03,mittermaier04,goehlert05}. Moreover,
it has been suggested that these changes, although small, may
correspond to entropy differences that, when converted into free
energies, are of the same order of same magnitude as the changes in
native state stability. Further, it has recently been suggested that
arginine to lysine mutations may entropically stabilize the native
state of hyperthermophilic proteins \cite{berezovsky05}.
Consequently, these changes in native state entropy may be important
factors to be considered in order to understand and to predict
stability changes upon mutation. A detailed analysis of the basis
for these changes is therefore of considerable significance.

Previous estimates of the free energy changes from order parameters
have made two assumptions about the relationship between measured
order parameters and native state entropies: (i)~that the
re-orientations of individual atomic bond vectors are independent of
each other, and (ii)~that the motions involved can be described by
models that are sufficiently simple that an analytical relationship
can be derived between order parameters and entropies. Although both
computational and experimental studies
\cite{yang96a,prompers00,stone01a,lee02a,mayer03} suggest that these
approximations are relatively good, in particular when calculating
entropy changes, it still remains important to test their validity.
The method of analysing changes in native state entropy that we have
here presented does not require these approximations. The use of MD
simulations restrained by $S^2$ values enables us to replace the
analytical models (e.g. that described as `diffusion in a cone')
with a procedure in which the underlying CHARMM molecular mechanics
force-field is used to generate a detailed description of the
possible intramolecular motions. Using this procedure we find that
the experimental changes in order parameters in the F20L and F20V
variants of Fyn SH3 translate into free energy changes on the order
of several kcal mol$^{-1}$. These values are of the same order of
magnitude as the stability changes that accompany these mutations
and therefore constitute an important term in the overall effect on
stability.

The procedure that we have used for estimating entropy changes is
based on a set of assumptions that differs from those discussed
above and include the use of (i) restrained MD simulations to sample
the conformational space, and (ii) a quasi-harmonic approximation to
describe the molecular motions. Since these assumptions are
different from those in the semi-analytical approach, it is highly
encouraging that the two methods give estimates of entropy changes
of the same sign and magnitude. A more detailed analysis will
require the development of methods for estimating entropies from
structural ensembles that take anharmonic contributions more fully
into account \cite{cheluvaraja05}. Finally, both procedures used to
relate the entropy changes to the variations in the free energy of
unfolding assume that the changes in vibrational entropy of the
unfolded state are negligible. This assumption may hold for a highly
denatured state in which specific contacts are absent, although in
some proteins amino acid mutations are known to cause long range
changes in the structure and dynamics of the unfolded state
\cite{klein-seetharaman02,fieber04}. The thermodynamic consequences
of such changes is, however, difficult to estimate due to the
theoretical problems in relating relaxation experiments for
disordered systems to structural features and due to sampling
problems for unfolded proteins in molecular dynamics simulations.

Proteins are stabilized by a multitude of weak interactions, and
protein structures can be considered at least partially liquid-like
\cite{zhou99,lindorff05a,best05}. In such systems the removal or
weakening of stabilizing interactions is expected to cause an
increase in the fluctuations in the remainder of the system, and
hence to increase its configurational entropy
\cite{dunitz95,qian98}. Experimentally, this effect has been
difficult to observe in relation to protein engineering experiments.
This is partially due to technical issues pertaining to the methods
by which entropy and enthalpy is measured \cite{beasley02}, as well
as the large solvent contributions to thermodynamic parameters
\cite{vaughan99}. The procedure that we here describe allows for a
direct and quantitative determination of changes in the
configurational entropy of proteins, and thus provides a crucial
tool to understand the complicated interplay between stabilizing
interactions and molecular fluctuations.

The observation that the experimentally observed decrease in
stability of the native states of the F20L and F20V mutant proteins
may include large contributions from changes in native state
configurational entropy that oppose a loss in enthalpy has important
consequences for the understanding and use of the protein
engineering method to modulate stabilities. Thus, models that
rationalize the stability change of mutant proteins in terms of
changes in intramolecular contacts and solvation energies should
also include changes in the residual entropy of the native states.

Finally, the increasing availability of experimental measurements
of backbone and side-chain dynamics is making it possible to predict order
parameters from protein structures with reasonable accuracy
\cite{goodman00,zhang02,ming04}. If these types of procedures
become sufficiently accurate to enable reliable predictions
to be made of the changes in order parameters
from mutations it should be possible to calculate the corresponding
changes in residual entropies using the approach described here
without the need of experimental measurements.
The resulting values can then be used as an integral
part of design procedures that
attempt to modulate protein stability by means of amino acid substitutions.
%
\section*{Acknowledgments} KLL was supported by the Danish Research
Agency and an EMBO long-term fellowship. LEK acknowledges support
from the Canadian Institutes of Health Research. LEK holds a Canada
Research Chair in Biochemistry. MV is a Royal Society University
Research Fellow. The research of CMD and MV is supported in part by
Programme Grants from the Wellcome and Leverhulme Trusts.
%
\newpage
\bibliographystyle{jmb}
\bibliography{bib-database}
\newpage
\section*{Captions} %
\begin{description}
\item[Fig.~\ref{fig:structures}] Structural ensembles generated by
        restrained molecular dynamics simulations.
        Simulations were carried out for the wild-type (WT) Fyn SH3 domain, as
        well as two mutant forms (F20L and F20V), using
        $S^2$ values for amide and methyl mobilities as restraints \cite{best04}.
\item[Fig.~\ref{fig:order-parameters-delta}]
Back-calculation of $S^2$ values from structural ensembles generated
by the restrained molecular dynamics approach. {\textbf{A}}: $S^2$
values for both amide and methyl groups were back-calculated and
compared to the experimental values for the WT Fyn SH3 domain and
its two mutational variants (F20L and F20V). {\textbf{B}}:
Correlation between experimental and calculated changes in order
parameters ($\Delta(S^2)=S^2_{mut} - S^2_{WT}$) for both the F20L
and the F20V mutants of the Fyn SH3 domain.
\item[Fig.~\ref{fig:dihedrals}]
Distributions of side-chain $\chi$ dihedral angles \cite{markley98}
for 10 side-chains in the WT Fyn SH3 and its two mutational variants
F20L and F20V. In each plot is shown the experimentally observed
$\Delta(S^2)=S^2_{mut} - S^2_{WT}$ values for both F20L (blue) and
F20V (red). The $\Delta (S^2)$ values shown for Leu29 and Val55
residues are the average values for the $\delta$ and $\gamma$ methyl
groups, respectively. The histograms show that the experimentally
observed changes in $S^2$ values for methyl group mobility can be
explained by relatively small variations in the distributions of
side-chain dihedral angles.
\item[Fig.~\ref{fig:quasi}] Estimating the entropic effects of the
    mutations using a quasi-harmonic analysis.
    {\textbf{A}}: Cumulative density of states for
    vibrational modes in WT, F20L and F20V Fyn SH3. The frequencies
    were obtained from a quasi-harmonic analysis \cite{brooks95} of the
    structural ensembles as described in the methods section. The insert
    shows an expansion of the region between 0--2000 ns$^{-1}$. In
    the frequency range larger than 2000 ns$^{-1}$ the lines are
    essentially parallel meaning that the vibrational differences
    are found in the low frequency range.
    {\textbf{B}}: The configurational entropy was estimated from the
    vibrational frequencies and was used to calculate
    $T\Delta S_{vib}$ for the two mutations. The figure shows the
    results when different number of modes (starting from the low
    frequency end) are included in the calculations.
    Error bars correspond to one
    standard deviation and were estimated using a bootstrap
    procedure. The larger error bars obtained when including more
    than $\approx250$ modes in the analysis (corresponding to
    frequencies larger than $\approx1000\textrm{ns}^{-1}$), is caused by
    the fact that bond lengths were kept approximately constant
    during the simulations \cite{ryckaert77,best04}.
\end{description}
\newpage
\section*{Figures} %
\newpage
\newpage
\newpage
\end{document}